\documentclass{article}

\usepackage{arxiv}

\usepackage[utf8]{inputenc} % allow utf-8 input
\usepackage[T1]{fontenc}    % use 8-bit T1 fonts
\usepackage{hyperref}       % hyperlinks
\usepackage{url}            % simple URL typesetting
\usepackage{booktabs}       % professional-quality tables
\usepackage{amsfonts}       % blackboard math symbols
\usepackage{nicefrac}       % compact symbols for 1/2, etc.
\usepackage{microtype}      % microtypography
\usepackage{graphicx}
\usepackage{amsmath}
\usepackage{amssymb}
\usepackage{lineno}
\usepackage{hyperref}
\usepackage{listings}

\title{PyLops - A Linear-Operator Python Library for large scale optimization}

\author{
  Matteo Ravasi \\
  Equinor ASA\\
  Bergen, Norway \\
  \texttt{matteoravasi@gmail.com} \\
   \And
 Ivan Vasconcelos \\
  Department of Earth Sciences\\
  Utrecht University\\
  Utrecht, The Netherlands \\
  \texttt{i.vasconcelos@uu.nl}}
  
\begin{document}
\maketitle

\begin{abstract}
Linear operators and optimisation are at the core of many algorithms used in signal and image processing, remote sensing, and inverse problems. For small to medium-scale problems, existing software packages (e.g., MATLAB, Python numpy and scipy) allow for explicitly building dense (or sparse) matrices and performing algebraic operations (e.g., computation of matrix-vector products and manipulation of matrices) with syntax that closely represents their corresponding analytical forms.
However, many real-application, large-scale operators do not lend themselves to explicit matrix representations, usually forcing practitioners to forego of the convenient linear-algebra syntax available for their explicit-matrix counterparts. PyLops is an open-source Python library providing a flexible and scalable framework for the creation and combination of so-called \textit{linear operators}, class-based entities that represent matrices and inherit their associated syntax convenience, but do not rely on the creation of explicit matrices.
We show that PyLops operators can dramatically reduce the memory load and CPU computations compared to explicit-matrix calculations, while still allowing users to seamlessly use their existing knowledge of compact matrix-based syntax that scales to any problem size because no explicit matrices are required.
\end{abstract}

%\keywords{Linear algebra \and Inverse problems \and Optimization \and Linear operator \and Software \and Python}

%% main text
\section{Introduction}
\label{intro}

Numerical linear algebra is at the core of many problems in signal processing \cite{kay1993}, image processing \cite{gonzalez2017}, inverse problems \cite{hansen2010, bertero1998} with applications to remote sensing \cite{twomey1997}, geophysics \cite{claerbout2014}, medical imaging \cite{suetens2009}, and even some areas of machine learning such as deep neural networks \cite{goodfellow2017}. 

Commonly used within these disciplines is the notion of \textit{linear operator}, mapping vectors from one space (generally referred to as the model space) into another space (referred to as the data or observation space), and \textit{inverse problem}, which is the process of estimating from a set of observations the causal factors that produced them, or in other words the underlying model vector \cite{hansen2010}. 

Three alternative approaches can be identified for solving an inverse problem: explicit solvers with dense matrices, iterative solvers with dense matrices, iterative solvers with linear operators. For problems of fairly limited size and/or when the computation power at hand allows it, one can first create a dense (or sparse) matrix and subsequently exploit the power of explicit solvers or analytical pseudo-inverse formulas to directly invert such a matrix. Finally, the inverse matrix is multiplied to the observation vector to obtain an estimate of the model. This route is however not always viable and iterative solvers, most belonging to the family of gradient-descent methods, are usually employed in real-life applications. A clear advantage of such family of solvers is that one does not need direct access to the matrix, rather it only needs to be able to compute the forward and adjoint operations. Several software packages provide core functionalities for dealing with arrays and matrices, as well as a suite of explicit and iterative numerical solvers: this is for example the case of MATLAB, the Python scientific libraries numpy and scipy, as well as the more low level libraries BLAS and LAPACK. Moreover, several open-source projects provide high-level, easy-to-use  routines for the numerical treatment of ill-posed problems, such as the Regularization Tools package \cite{hansen2007}. 

The widespread need to perform core linear algebra operations as efficiently and fast as possible has led to large computing technology investments in the last two decades, focused on the search for efficient implementations on CPUs (using both multi-threading as well as multi-core paradigms), GPUs, and more recently TPUs. An example of a field that has massively benefited from such advances is that of machine learning, and more specifically \textit{deep learning}: very complex, deep neural networks with millions of weights (i.e., model parameters) can in fact solved today in a matter of hours mostly because of the incremental gain in speed in matrix-matrix and matrix-vector computations that GPUs (or TPUs) provide when compared to CPUs. Frameworks such as Theano \cite{theano2016}, tensorflow \cite{tensorflow2015} or PyTorch \cite{paszke2017} have been developed to specifically satisfy such a need and take advantage of advances in hardware components. 

Nevertheless, when dealing with a large variety of physics-based inverse problems, the underlying linear operators are often far from being dense matrices (as opposed to, for example, dense layers in a neural network). Instead, they can be represented via sparse, structured matrices with fewer non-null elements compared to the null ones. Linear operators take advantage of such a structure in order to produce computer code for the application of forward and adjoint operations that is efficient and scales with problem size. Such computer code can be written in a manner that inherits the syntax convenience of analytical linear algebra, by simply representing forward and adjoint operations via class-defined methods that reproduce the result of otherwise explicit matrix-vector (or matrix-matrix) products. This construct not only serves both sparse (e.g. physics-based) and dense operators (e.g., convolution with Green's functions or neural-network layers) equally well, but it also provides full functionality for the use of iterative solvers. Conveniently, the Python library \texttt{scipy} provides a barebone, generic class for the definition and application of linear operators, which we leverage from and build on within the PyLops package as discussed below. Other examples of currently available software packages that provide a general interface to linear operators are the C++ \texttt{Rice Vector Library} \cite{padula2009}, the MATLAB \texttt{Spot} software package \cite{spot} and the Python \texttt{fastmat} library \cite{wagner2017}. Moreover, some open-source software packages that employ a similar construct to solve domain-specific large-scale linear inverse problems are the \texttt{ASTRA-toolbox} \cite{vanaarle2015}, \texttt{Seplib} \cite{clapp2012, claerbout2014}, and \texttt{Madagascar} \cite{m8r}. Many of the packages mentioned above however tend to prioritize the ability of solving large inverse problems efficiently in exchange for a loss in the convenient linear-algebra syntax. To the best of our knowledge, only the \texttt{Spot} package, and to a lesser degree the Python \texttt{fastmat} library, achieve the best of both worlds. PyLops is a Python library that accomplishes the very same goal while at the same time being more tightly connected to the Python ecosystem by directly building on top of the linear operator definition within the \texttt{scipy} library.

\section{A brief tour of linear operators}
\label{problem}
A linear operator can be formally represented as a matrix-vector (or matrix-matrix) multiplication: 

\begin{equation}
\label{eq:linearity}
\mathbf{y} = \mathbf{A} \mathbf{x}
\end{equation}

where $\mathbf{A}\in \mathbb{R}^{(N \times M)}$ (or $\mathbb{C}^{(N \times M)}$) is an operator that maps a model vector $\mathbf{x}$ belonging to the real space $\mathbb{R}^M$ (or complex space $\mathbb{C}^M$) into a data vector $\mathbf{y}$ belonging to the real space $\mathbb{R}^N$ (or complex space $\mathbb{C}^N$). The linear mapping from a known set of input parameters ($\mathbf{x}$) into a vector in the data space ($\mathbf{y}$) is generally referred to as \textit{modelling} or the \textit{forward problem}. 

As already mentioned, several linear mappings tend to obey to a certain structure and exploiting such a structure when applying them to a vector can usually lead to a significant gain in both computation speed and memory efficiency. One common example are operators that can be expressed in terms of a convolution (correlation) between the model (data) vector and a compact kernel. Operators of such a kind can be implemented by creating a Toeplitz matrix that contains the elements of the kernel, followed by a matrix-vector multiplication with the model (or data) vector. When the kernel is compact, such matrix is a very sparse, band matrix with few non-zero elements around the main diagonal and zeros elsewhere. Performing the matrix-vector multiplication leads to poor performance, as several multiplications and summations with zero elements are performed. For example, imagine we want to apply a first-order derivative to a vector $\textbf{x}$. The first-order derivative, in its simplest form, can be approximated by a two-sample stencil $\lbrack 1/dx, -1/dx \rbrack$ applied to each pair of samples of the input vector; as for any convolutional operator with a generic kernel, the very same operation can be performed in different ways:
\begin{enumerate}
    \item create a dense matrix with $1/dx$ along the main diagonal and $-1/dx$ along the first lower diagonal (and zero elsewhere), followed by a matrix-vector multiplication,
    \item convolve the input signal by the stencil,
    \item subtract each sample of the input vector by the previous sample (i.e., $y_i = (x_{i+1}-x_i)/dx$).
\end{enumerate}
This very last approach is the one adopted in the PyLops implementation of a first-order derivative as it does not only remove the need for storing $-1/dx$ and $1/dx$ values, but it also reduces the number of operations to two multiplications and one summation for each sample of the output vector. More in general, PyLops philosophy is to devise ad-hoc implementations for different operators with the aim of exploiting their specific structure and reduce both memory usage and computational cost.

An additional benefit of using linear operators becomes evident when attempting to solve an inverse problem. Without loss in generality, we consider the least-squares solution to an over-determined inverse problem ($n>m$):

\begin{equation}
\label{eq:inverse}
\hat{\mathbf{x}} = \min_{\mathbf{x}} || \mathbf{y} - \mathbf{A} \mathbf{x} ||_2  \rightarrow  \hat{\mathbf{x}} = (\mathbf{A}^H \mathbf{A})^{-1} \mathbf{A}^H \mathbf{y}
\end{equation}

\noindent Notice how the solution $\hat{\mathbf{x}}$ that minimizes the cost function requires both the operator $\textbf{A}$ and its adjoint $\textbf{A}^H$. This is not only the case when an explicit solutions is used (least-squares in this case), but also when solving the problem by means of, e.g., an iterative gradient-based solver~\cite{hansen2010}. Thus, working with explicit matrices requires creating and storing also the adjoint matrix, doubling the amount of data in memory. On the other hand, linear operators can implement the adjoint in a similar fashion as the forward by exploiting the structure of the operator itself, leading to limited or (in most cases) no additional storage being required. %In the example of a first-order derivative operator, the adjoint is in fact simply obtained by summing shifted versions of the input vector belonging to the data space, i.e. $x_i = x_i-y_i/dx$ and $x_{i+1} = x_{i+1}+y_i/dx$).

\section{Software Framework}
\label{software}
PyLops' main goal is to provide an easy-to-use Application Programming Interface (API) to create and solve inverse problems by means of linear operators and express them in a way that mimics as closely as possible the analytical linear algebra formalism used to describe the problem in the first place. Moreover, the library efficiently scales to problems of any size, as shown in the benchmarking tests in section 5. To achieve these goal, each linear operator is a class-based entity and different operators can be both used independently, combined together by means of basic mathematical operations (e.g., $+$, $-$, $*$ - see below for more details), and fed directly into various solvers. Taking a modular approach to the creation of linear operators, the library makes it easy for other developers to implement new linear operators and to seamlessly include them in the framework. This ultimately enables the combination of any new and existing operators, providing an easy and quick way to experiment with novel inverse problems.

\begin{figure}[htb]
\centering
  \includegraphics[scale=0.3]{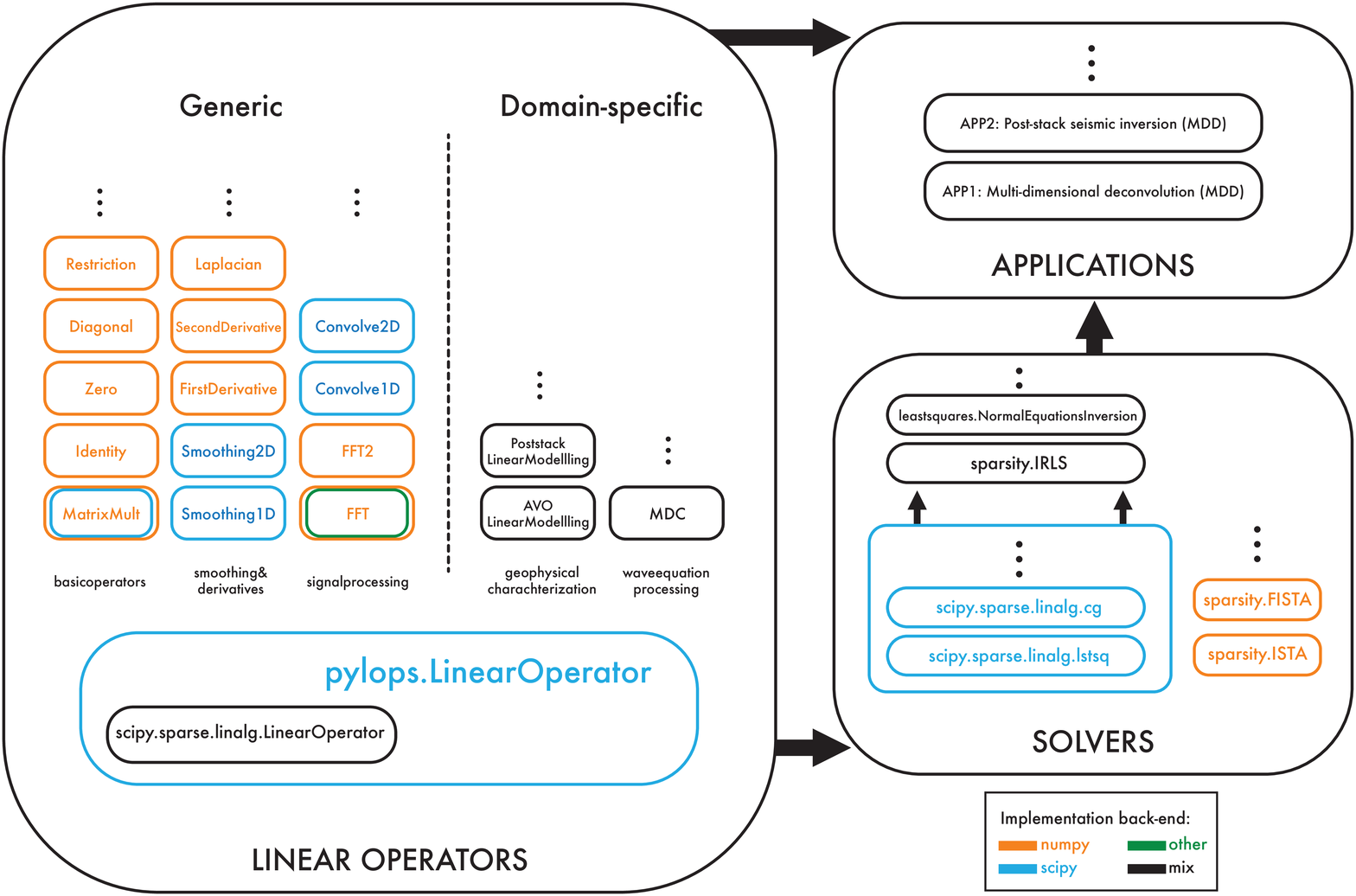}
  \caption{Schematic representation of the software API. Colors  indicate which library is used in the back-end of a linear operator (or solver).}
  \label{fig:architecture}
\end{figure}

The API can be loosely seen as composed of three inter-connected units as shown in Figure \ref{fig:architecture}.

\subsection{Linear Operators}
\label{linearop}
The first unit contains the entire suite of linear operators. \texttt{pylops.LinearOperator} is the main class of the library which is used as parent class for all other linear operators, such that they inherit its various internal methods as described below. Submodules are used to create an organized stack of operators and separate basic operators (that are used within several applications) to more domain-specific ones, as, e.g., within the \texttt{signalprocessing} submodule.

\texttt{pylops.LinearOperator} creates a generic interface for matrix-vector (and matrix-matrix) products that can ultimately be used to solve any forward or inverse problem of the form $\mathbf{y} = \mathbf{A} \mathbf{x}$. This is achieved by overloading the \textit{scipy} class \texttt{scipy.sparse.linalg.LinearOperator}, on top of which additional properties and methods are also defined. Forward and adjoint matrix-vector operations are achieved by implementing the method \texttt{\_matvec} for the forward, and \texttt{\_rmatvec} for the adjoint. The attributes \texttt{shape} (tuple of two integers) and \texttt{dtype} must also be provided during initialization to identify the shape and type of the operator itself. Moreover, \texttt{pylops.LinearOperator} requires an additional boolean attribute \texttt{explicit}, which identifies whether the operator has an explicit or implicit matrix representation. This allows to infer the most appropriate solver to be used when invoking the  \texttt{\_\_truediv\_\_} method as explained below.

Any PyLops linear operator is mathematically speaking an object that is both concept- and syntax-wise equivalent to a matrix. As such, enabling user to combine those operators actually reduces to implementing the following five elementary operations: sum, multiply by a scalar, multiply (or chain) operators, stack vertically and stack horizontally. With the aim of being able to write code that resembles as much as possible the underlying mathematical equations, we take advantage of the ability of Python to perform \textit{operator overloading} of various magic methods --- i.e., methods with the double underscores at the beginning and the end - to allow using mathematical symbols such as $+$, $-$, $*$, and $/$ to perform those elementary operations. More specifically, the following operator overloads are implemented: 

\begin{itemize}
\item \texttt{\_\_matmul\_\_} (called via $@$) or \texttt{\_\_mul\_\_} (*): when applied to a numpy \texttt{ndarray} vector, executes the forward computation of matrix-matrix or matrix-vector multiplication, respectively;
\item \texttt{\_\_mul\_\_} ($*$): when applied to a scalar, left- or right-multiplies the operator by a scalar, while when applied to another \texttt{LinearOperator}, chains the two operators;
\item \texttt{\_\_add\_\_} ($+$): when applied to another \texttt{LinearOperator}, sums the two operators;
\item \texttt{H} and \texttt{T}: creates the transpose (or hermitian operator) and performs the adjoint computation when combined with a \texttt{*} (i.e., \texttt{.H*});
\item \texttt{\_\_truediv\_\_} ($\backslash$): when applied to a numpy \texttt{ndarray} vector, solves the inverse problem $\mathbf{y} = \mathbf{A} \mathbf{x}$ with either explicit or iterative solver.
\end{itemize}

Additionally, two other convenience methods are implemented within the \texttt{pylops.LinearOperator} class:

\begin{itemize}
\item \texttt{eigs}: estimate the singular values of the operator using the scipy wrapper of ARPACK fortran package \cite{arpack}.
\item \texttt{cond}: use the \texttt{eigs} method to compute the conditioning number (the ratio of the largest-to-smallest eigenvalues).
\end{itemize}

\subsection{Solvers}
\label{solvers}
Solving a linear problem by means of an off-the-shelf least-squares cost function as in equation \ref{eq:linearity} may not always provide a good estimate of the inpu model~\cite{hansen2010}; this is always the case for ill-posed linear operators that cannot be inverted directly (e.g., the \texttt{Restriction} operator, as shown in the numerical example in section 4) or in the presence of noisy data. In order to obtain an improved estimate of the input model, regularization terms can be included in the cost function (i.e., adding terms such as the well-known Tikhonov regularization $|| \mathbf{x}||_2$ or sparsity promoting terms such as $||\mathbf{x}||_1$) and/or the model can be preconditioned (i.e., solving for a preconditioned model $\mathbf{p}$ such that  $\mathbf{x} = \mathbf{P}\mathbf{p}$ where $\mathbf{P}$ could be a smoothing operator). While large variety of linear solvers (e.g., conjugate-gradient solver \cite{hestenes1952} or the LSQR solver \cite{paige1982}) is currently available in the public domain, for example as part of the \texttt{scipy} package, and more specifically the \texttt{scipy.sparse.linalg} submodule within the Python ecosystem, the user is generally left with the task of adding regularization and/or preconditioning terms. PyLops provides thin wrappers around some of those solvers and eases the use of regularization and/or preconditioning in inverse problems with very little amount of extra code. Our entire suite of \textit{enriched} solvers is provided in the submodule \texttt{pylops.optimization} and subdivided into least-squares within \texttt{pylops.optimization.leastsquares} and sparsity-promiting solvers within \texttt{pylops.optimization.sparsity}.

\subsection{Software Dependencies}
\label{software_dependencies}
PyLops relies and builds on top of the two main external libraries for scientific computing in Python, \texttt{numpy} \cite{numpy} and \texttt{scipy} \cite{scipy2001}, for all its linear operators and solvers. 

In some circumstances, additional optional back-ends are also implemented to improve the performance of forward and adjoint operations. This is for example the case of the \texttt{FFT} operator, where a fast implementation of the Fast Fourier Transform (FFT) algorithm is provided by the library \texttt{pyfftw}, which is a python wrapper around the famous FFTW library \cite{frigo1998}. In this case PyLops provides two back-end options (referred in the code as \texttt{engine}), one using numpy's implementation of fft and ifft, and another using \texttt{pyfftw}. In the case a user uses \texttt{engine='fftw'} whist not having \texttt{pyfftw} and \texttt{FTTW} installed, PyLops automatically falls back to the numpy implementation. A similar approach is taken also for the \texttt{Radon2D} operator, where \texttt{numba} is used in this case to speed-up for loops computations: again, a fallback numpy engine is implemented to keep \texttt{numba} as an optional dependency. 

\subsection{Testing and operator validation}
\label{testing}
In the framework of linear operators, a very strong indication of the correctness of the forward and adjoint implementations is the so-called \textit{dot-test}. More specifically, two random vectors $\textbf{u}$ and $\textbf{v}$ of size $\lbrack M \times 1 \rbrack$ and $\lbrack N \times 1 \rbrack$ are generated, forward and adjoint operations performed as in equation \ref{eq:dottest}, and the following equality tested within a certain tolerance:
\begin{equation}
\label{eq:dottest}
(\mathbf{Op}*\mathbf{u})^H*\mathbf{v} = \mathbf{u}^H*(\mathbf{Op}^H*\mathbf{v})
\end{equation}

Alongside with the dot-test, we also solve a small-scale inverse problem for every linear operator. The inverted model is compared to the original one used to model the data and it is checked that the two vectors match within a certain tolerance. It is important to remember that some inverse problems, especially those with an under-determined operator ($N < M$), do not always have a unique solution and a satisfactory inverted model can only be obtained by including additional prior information in the form of additional regularization (as shown in one of the examples below).

Tests have been implemented using \texttt{pytest} and are connected to a continuous integration (CI-Travis and Azure Pipelines) system such that new pull requests are safeguarded. Automated tests cover all the linear operators, and multiple tests have been implemented to validate different combinations of (both mandatory and optional) input parameters. At the time of writing, PyLops has over 300 automated test with a code coverage of 86\% (estimate provided by Codacy).

\subsection{Contributing to the Software}
\label{software_contribution}
We foresee contributions from across different areas of scientific computing where inverse problems are applicable. In order to facilitate contributions we have created a check-list of four mandatory steps that are required for a new operator (or solver) to be accepted to become part of the codebase of PyLops - refer to $pylops.readthedocs.io/en/latest/adding.html$ for more details. By strictly adhering to these requirements, we strive to keep a well-maintained, well-tested, and well-documented codebase, while strongly encouraging external contributions.

\section{Code examples}
\label{examples}

Several tutorials are available as part of the documentation of PyLops, written using Sphinx-Gallery and hosted at \url{pylops.readthedocs.io}. Moreover, more in-depth code examples can be found at $github.com/mrava87/pylops\_notebooks$.

In this section we combine several linear operators and solvers with the aim of interpolating a one dimensional signal composed of three sinusoids and sampled at irregularly and coarsely spaced locations along the time axis.

\subsection{Sample code snippets for basic operators}
\label{samplecode}
First of all, we see how we can create and apply a \texttt{Restriction} operator $\textbf{R}$ within the PyLops framework. A restriction operator extracts a subset of $N$ values at locations \texttt{iava} from an input (or model) vector $\textbf{x}$ in forward mode:

\begin{equation}\label{eq:restrictionforward}
y_i = x_{l_i}  \quad \forall i=1,2,...,M
\end{equation}

where $\mathbf{l}=[l_1, l_2, l_M]$ is a vector containing the indices at which samples are taken from the original array. In adjoint mode, the values in the data vector $\mathbf{y}$ are placed at locations $\mathbf{L}$ in the model vector:

\begin{equation}\label{eq:restrictionadjoint}
 x_{l_i} = y_i  \quad \forall i=1,2,...,M
\end{equation}

and $x_{j}=0 \quad \forall j=1,2,,...N \quad (j \neq l_i)$ (i.e., at all other locations in input vector). 

\begin{figure}
  \centering
  \includegraphics[scale=.4]{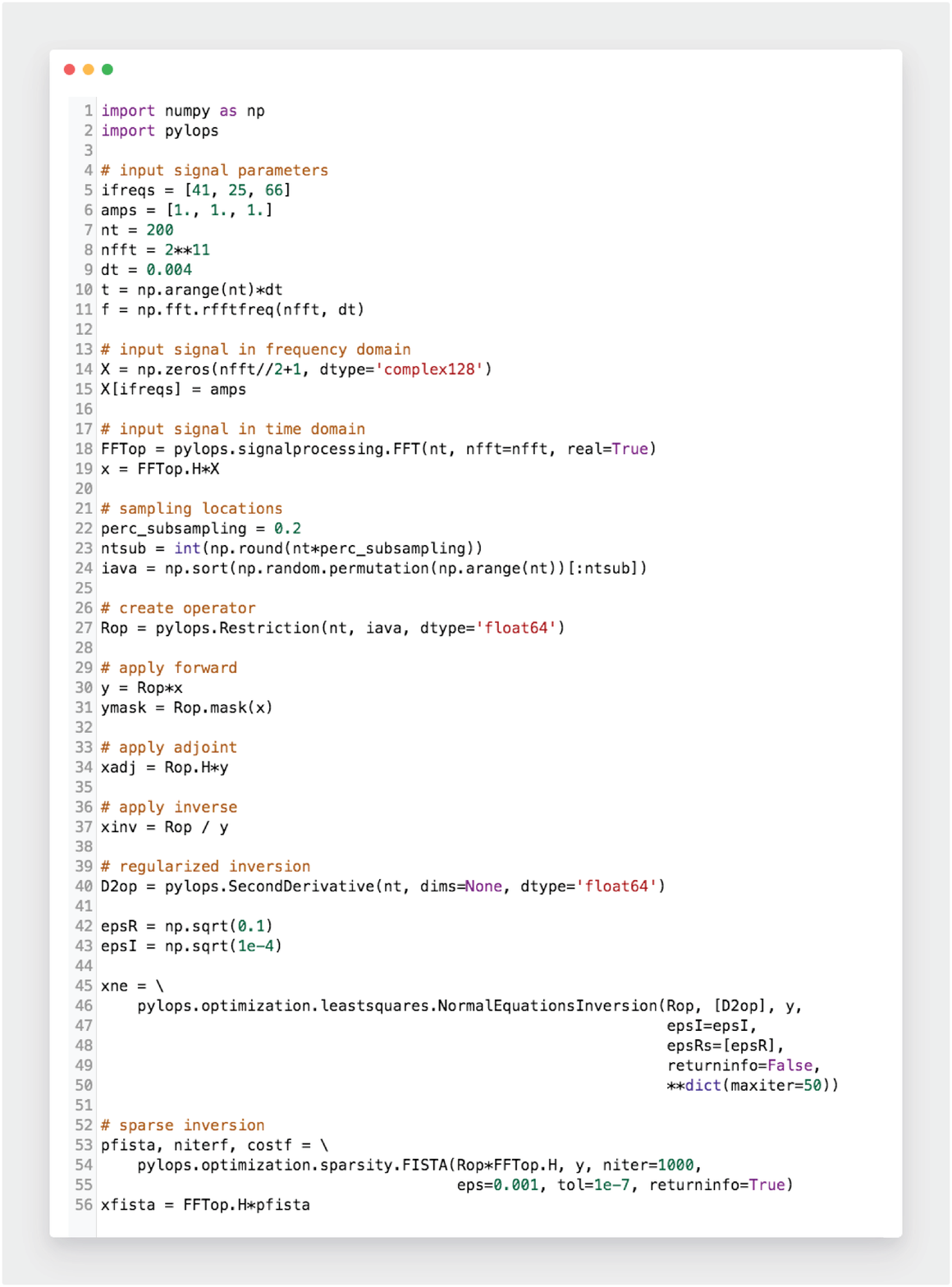}
  \caption{Code snippet for creation and application of forward, adjoint and inverse \texttt{Restriction} operator to a vector (created with CodeZen).}
  \label{fig:code}
\end{figure}

We can translate the following problem into computer code using PyLops, as shown in the code snippet in Figure \ref{fig:code}.

In this code snippet, we create an input signal composed of three sinusoids in the frequency domain (lines 5 -15), convert it to time domain using the \texttt{FFT} linear operator (18-19), define indices for sampling the signal at irregular locations (22-24), create the \texttt{Restriction} operator (27), apply it in forward mode to the input signal (30-31),  apply its adjoint to the calculated data (34), and finally invert the operator (37). 

\begin{figure}[htb]
\centering
  \includegraphics[scale=0.55]{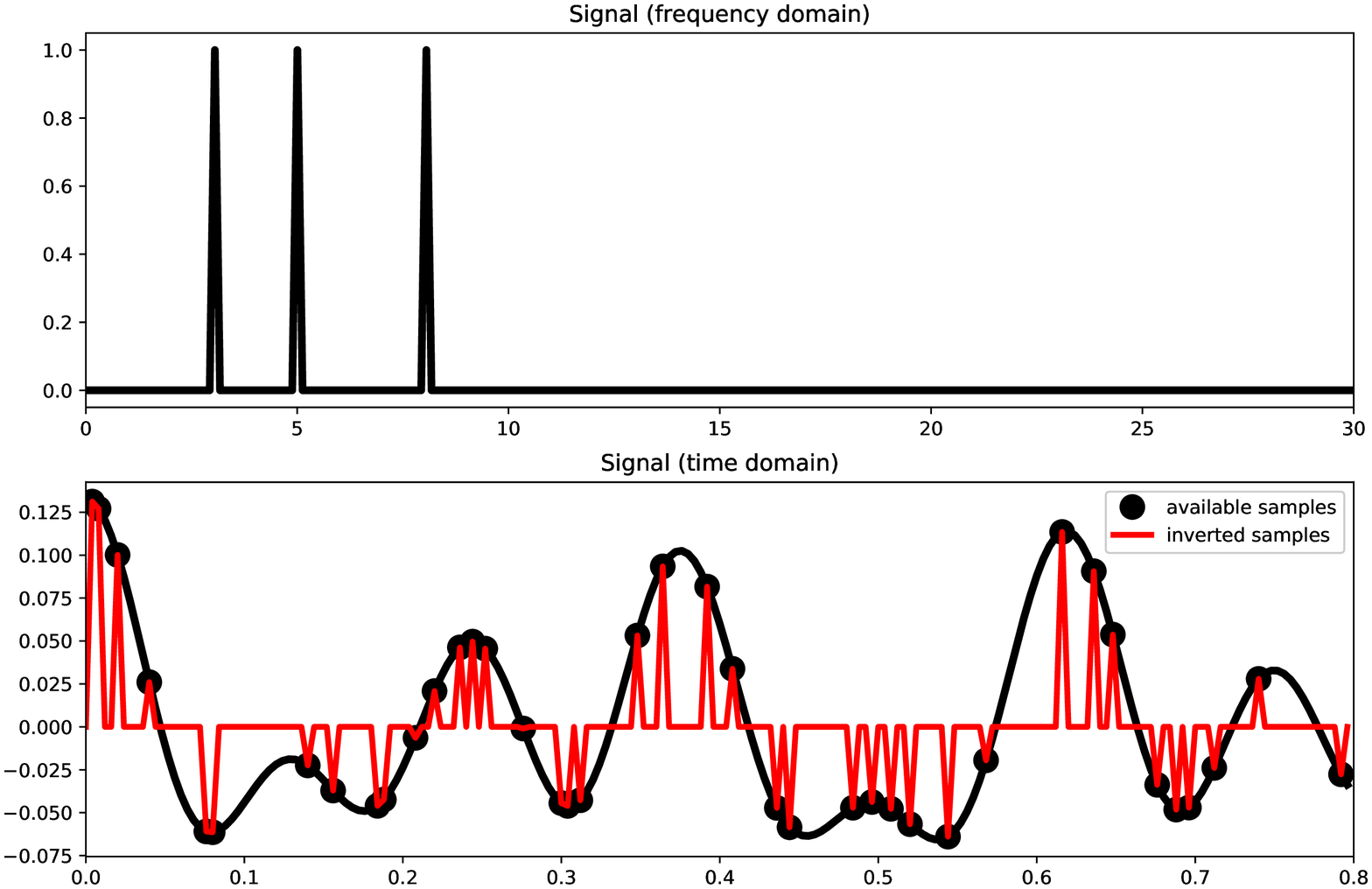}
  \caption{Input signal in (a) frequency domain \texttt{X} and (b) time domain \texttt{x}. Sampled signal (\texttt{y} - green dots) and inverted signal (\texttt{xinv} - red) are also shown in panel (b).}
  \label{fig:restriction_input}
\end{figure}
Figure \ref{fig:restriction_input} shows that the operator $\mathbf{R}$ is ill-posed and the inverse problem cannot be successfully solved by simply employing the magic method \texttt{/}. Such method does in fact implement the vanilla least-squares inversion (equation \ref{eq:linearity}) by means of the \texttt{scipy.sparse.linalg.lsqr} solver.

\begin{figure}[htb]
\centering
  \includegraphics[scale=0.55]{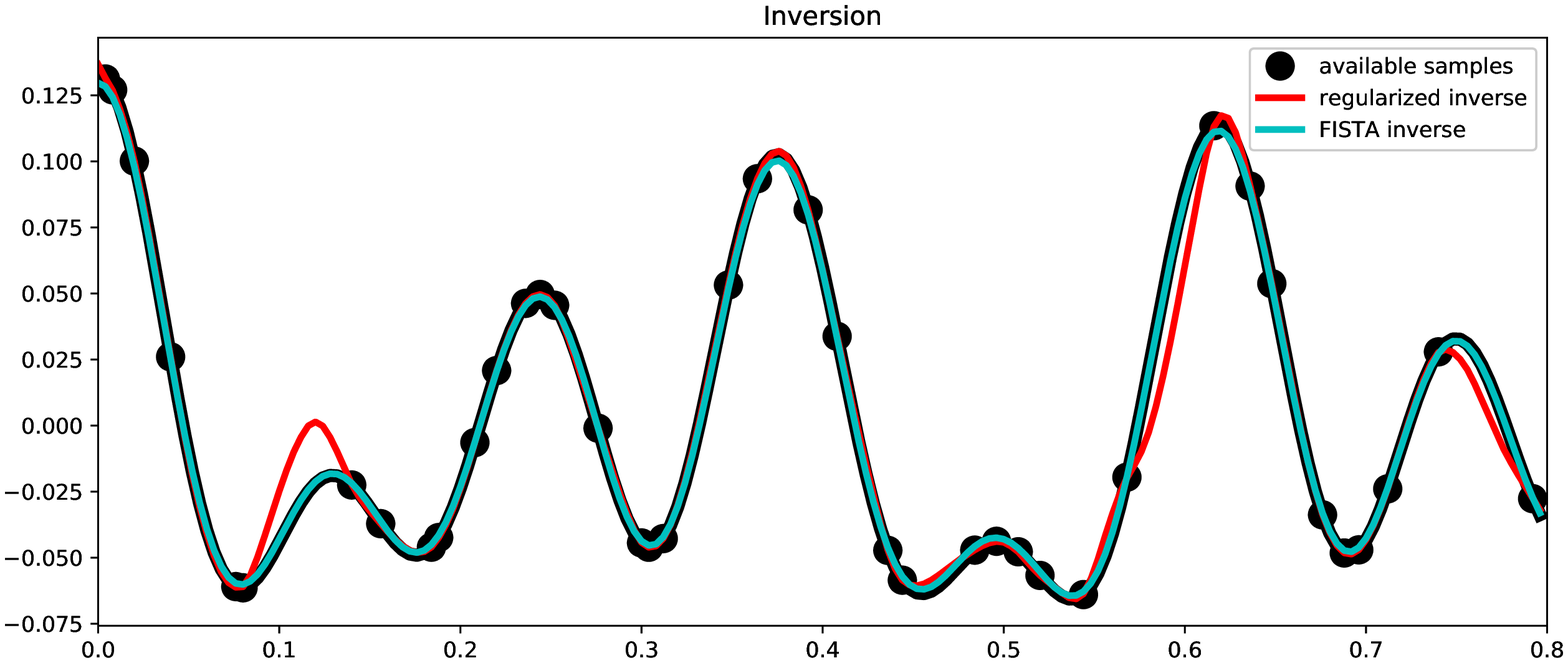}
  \caption{Recovered signal using regularized least-squares inversion (red) and sparsity-promoting inversion (cyan), and input signal (black).}
  \label{fig:restriction_inverse}
\end{figure}

In this example we show how we can improve our estimate by either i) including a regularization term that favours a smooth model by penalizing its second-order derivative ($\textbf{D} \textbf{x}$) or ii) taking advantage of the sparsity of the model in the frequency domain and using a sparsity promoting solver such as \texttt{pylops.optimization.sparsity.FISTA} \cite{beck2009}. As shown in figure \ref{fig:restriction_inverse}, the estimate of the input signal is very much improved in both cases.

\section{Benchmarking}
\label{benchmarking}
Finally, we analyze the three different linear operators used in the example from the standpoints of computational performance and memory usage. For each operator, we perform a benchmark test comparing the time it takes to apply the forward operator to an input vector using the PyLops implementation of such operator versus the application of a dot product with a dense (or sparse) matrix performing the very same operation. The comparison is done for operators of increasing size and the forward modelling is performed 200 times and logged via the Python \texttt{timeit.timeit} function. Comparisons are performed on a MacBook Air 1.3 GHz Intel Core i5 with a 8 GB 1600 MHz DDR3 RAM. Moreover, numpy and scipy are installed via the conda distribution and linked to the Intel MKL implementation of BLAS library for linear algebra. This leads to the best performance for the dot product on a CPU architecture as discussed in \cite{mklboost}.

\begin{figure}[htb]
\centering
  \includegraphics[scale=0.5]{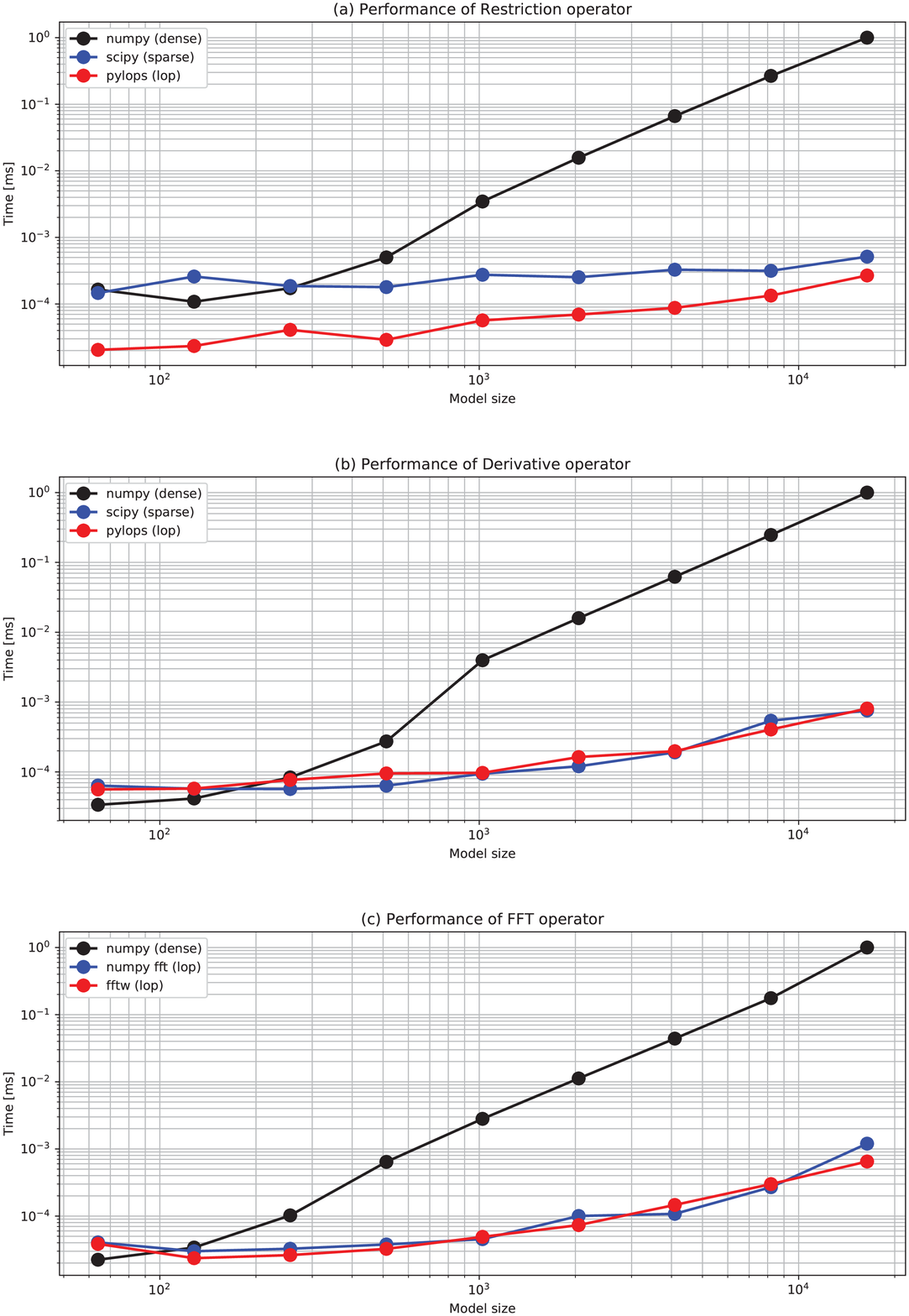}
  \caption{Performance benchmark of (a) \texttt{Restriction} operator, (b) \texttt{FirstDerivative} operator, and (c) \texttt{FFT} operator.}
  \label{fig:performance}
\end{figure}

For the \texttt{Restriction} operator (Figure \ref{fig:performance}a) we create both a dense matrix using \texttt{numpy.ndarray} and a sparse matrix using \texttt{scipy.sparse.csr\_matrix} as well as a PyLops operator. PyLops' implementation outperforms the na\"ive dot product for both dense or sparse matrices. Moreover, if we consider a model vector with $M=10^5$, and a subsampling of factor 10, the resulting data vector has size $N=10^4$. The dense matrix used to perform restriction of the model vector has a size of $N*M=10^9$ elements. Assuming each element to be a unsigned integer (8 bit), 8GB of memory is used to store this matrix (and another 8GB is required to store its adjoint). The memory usage is dramatically reduced for a sparse matrix, as three values need to be stored for each index where the input signal is sampled (row index, column index, and value - in this case 1); this amounts to about $3*M=30^4$ elements (960KB if we use int32 type for indices and values). A linear operator requires instead only storing the indices at which the input signal is sampled; in this case, that means only $N=10^4$ values (320KB if we use int32 type for the indices such values).

We consider now the two-point \texttt{FirstDerivative} operator (Figure \ref{fig:performance}b). This operator is convolutional in essence as it could be applied by convolving the input signal by a compact filter. The PyLops implementation outperforms the explicit dot product with a dense numpy matrix, while a similar performance is obtained in this case when a using a sparse-matrix. Though true for this isolated benchmark, we note that in real-applications multiple operators are generally chained (or stacked). Chaining explicit matrices will generally increase the complexity of the resulting matrix and \textit{densify} it, meaning that the resulting matrix is less sparse and the dot product less efficient. This is not the case for linear operators, where the computational time of a chained operator is equivalent to the sum of computational time of each operator. Morover, the memory usage for the \texttt{FirstDerivative} operator reduces to a single value, the step size $dx$, while for a dense matrix the size quadratically increases with the size of the model.

Lastly, we benchmark the Fast Fourier Transform \texttt{FFT} operator (Figure \ref{fig:performance}c). This is a peculiar case, as the FFT can be easily written as a fully-dense matrix and combined with other dense matrices as well as applied by means of a matrix-vector product. Using a linear operator we can however leverage from available open-source implementations of the FFT algorithm such as those in \texttt{numpy} or \texttt{FFTW} libraries. The storage in this case is also limited to a single number, the size of the FFT, while the required storage for the corresponding dense matrix increases again quadratically with the size of the model.

\section{Conclusions}
\label{conclusion}
In this paper, we present a general-purpose Python library for linear optimization, scaling from didactic numerical experiments to large-scale, real-life problems. Using the concept of \textit{linear operator classes} and taking advantage of several built-in functionalities of Python (e.g., operator overloading), a new Python framework is created whereby linear forward and inverse problems can be solved in a fully scalable manner (from tens to millions of model parameters). In addition to scalability, PyLops maintains, by design, a compact syntax that closely mimics the underlying analytical linear-algebra formulation of any chosen problem. Benchmark testing confirms that linear operators in PyLops scale well and efficiently with respect to more `na\"ive` implementations of the same operators by means of dense matrices. Moreover, the software architecture is created in a modular fashion, in such a way that it is very straightforward to create and include new linear operators (or solvers). Although not part of the current version of the project, the framework is not limited to linear inverse problems. PyLops could be used for solving nonlinear inverse problems of any kind, e.g., with optimisation methods that rely on linearised forward modelling, such as the widespread adjoint-state method. Moreover, there are ongoing activities in the following two sibling projects \textit{PyLops-GPU} and \textit{PyLops-distributed} to extend the capabilities of PyLops operators and solvers to GPU and distributed computing environments.

\section*{Acknowledgements}
\label{ack}
MR thanks Equinor for allowing the publication of this work. We also thank Joost van der Neut, Yanadet Sripanich, and Tristan van Leeuwen for insightful discussions. Jupyter notebooks are used to create Figures \ref{fig:restriction_input}, \ref{fig:restriction_inverse}, and \ref{fig:performance} can be found at $github.com/mrava87/pylops\_notebooks/$ $tree/master/papers/softwareX\_2019$. The authors cannot be held liable for any inappropriate use of this software library.

%% References: At least 5 are required 
%\bibliographystyle{unsrt}  
\bibliography{article}

\end{document}